\documentclass[journal]{IEEEtran}
\usepackage{cite}

\usepackage{amssymb}

\ifCLASSINFOpdf
  \usepackage[pdftex]{graphicx}
\else
  \usepackage[dvips]{graphicx}
\fi
\usepackage[cmex10]{amsmath}
\usepackage{array}

\begin{document}
%
\title{Brightness Control in Dynamic Range Constrained Visible Light OFDM Systems}
%
%
%


\author{Zhenhua~Yu,~\IEEEmembership{Student Member,~IEEE,}
        Robert~J.~Baxley,~\IEEEmembership{Member,~IEEE,}
        and~G.~Tong~Zhou,~\IEEEmembership{Fellow,~IEEE}
\thanks{Zhenhua Yu and G. Tong Zhou are with the School of Electrical and Computer Engineering, Georgia Institute of Technology, Atlanta,
GA 30332-0250, USA, e-mail: zhenhuayu@gatech.edu.}
\thanks{Robert J. Baxley is with the Georgia Tech Research Institute, Atlanta, GA 30332-0821, USA.}}

\maketitle

\begin{abstract}
Visible light communication (VLC) systems can provide illumination and communication simultaneously via light emitting diodes (LEDs). Orthogonal frequency division multiplexing (OFDM) waveforms transmitted in a VLC system will have high peak-to-average power ratios (PAPRs). Since the transmitting LED is dynamic-range limited, OFDM signal has to be scaled and biased to avoid nonlinear distortion. Brightness control is an essential feature for the illumination function. In this paper, we will analyze the performance of dynamic range constrained visible light OFDM systems with biasing adjustment and pulse width modulation (PWM) methods. We will investigate the trade-off between duty cycle and forward ratio of PWM and find the optimum forward ratio to maximize the achievable ergodic rates. 
\end{abstract}

\begin{IEEEkeywords}
Visible light communication (VLC), orthogonal frequency division multiplexing (OFDM), brightness control, pulse width modulation (PWM).
\end{IEEEkeywords}

%
\IEEEpeerreviewmaketitle

\section{Introduction}
\label{}
%
%
%
%
\IEEEPARstart{M}{otivated} by the rapid progress of solid state lighting technology and increasingly saturated radio frequency (RF) spectrum, visible light communication (VLC) has become a promising candidate to complement conventional RF communication \cite{o2008visible,elgala2011indoor}. VLC uses the visible light spectrum to transmit information.
In VLC, simple and low-cost intensity modulation and direct detection (IM/DD) techniques are employed, thus only signal intensity information, not phase information, is modulated. IM/DD requires the electric signal to be real-valued and unipolar (positive-valued). Recently, OFDM has been considered for VLC due to its ability to boost data rates and effectively combat inter-symbol-interference (ISI) \cite{hranilovic2005design,armstrong2009ofdm, Yu2012a}.  However, OFDM is known for its disadvantage of high peak-to-average power ratio (PAPR) and thus is very sensitive to nonlinear distortions. The LED is the main source of nonlinearity in VLC. Although LEDs can be linearized by a predistorter  \cite{Elgala2009}, the dynamic range is limited by the turn-on current and maximum permissible alternating current. A linear scaling and biasing model has been proposed in \cite{Yu2013} to make the OFDM signal work with the dynamic range constrained VLC system.

Brightness control is essential for the illumination function of VLC. Generally, there are two ways to control the brightness: (i) adjust the average input current; (ii) change the duty cycle of pulse width modulation (PWM). For single carrier pulsed modulation, the standard IEEE 802.15.7 \cite{Rajagopal2012} has applied the above two ways to control the brightness for on-off keying (OOK) and variable pulse position modulation (VPPM). A multiple pulse position modulation is proposed in \cite{Lee2011g}, which controls the brightness by changing the number of pulses in one symbol duration. Reference \cite{Choi2012} studied the optical power distribution, eye diagrams, and bit error rates of the PWM-based position modulation (PPM). A brightness control scheme was proposed in \cite{Baia} for overlapping PPM. Brightness control has also being investigated for the OFDM. In \cite{Ntogari2010}, PWM is combined with OFDM by sending the product of the OFDM and PWM waveforms. In reference \cite{Stefan2012}, the authors adjusted the average optical power of asymmetrically clipped optical OFDM (ACO-OFDM) and studied the nonlinearity of LED. In reference \cite{Wang}, the authors investigated the performance of M-QAM OFDM with PWM brightness control. To the best of our knowledge, however, only the reference \cite{Stefan2012} considered the dynamic range constraints. Moreover comparison between the average current adjusting method and the PWM method is still lacking. 

In this paper, we will analyze the performance of dynamic range constrained visible light OFDM systems with biasing adjustment method and PWM method. Biasing adjustment method sets the biasing level equal
to desired value for each OFDM symbol. PWM method can increase the signal power but with the expense of spectral efficiency. We will apply the two methods to DC biased optical OFDM (DCO-OFDM) \cite{hranilovic2005design} and compare their achievable ergodic rates. We will also jointly adjust the PWM biasing level and the PWM duty cycle and investigate their trade-off.

\section{System model}
\label{}
\subsection{Dynamic range constrained visible light OFDM system}
\label{}
In VLC systems, 
intensity modulation (IM) is employed at the transmitter. The forward signal $y(t)$ drives the LED which in turn converts the magnitude of the input electric signal $y(t)$ into optical intensity. The human eye cannot perceive fast-changing variations of the light intensity, and only responds to the average light intensity.  Direct detection (DD) is employed at the receiver. A photodiode (PD) transforms the received optical intensity into the amplitude of an electrical signal. 

\begin{figure}[!t]
  \centering
  \includegraphics[width=6cm]{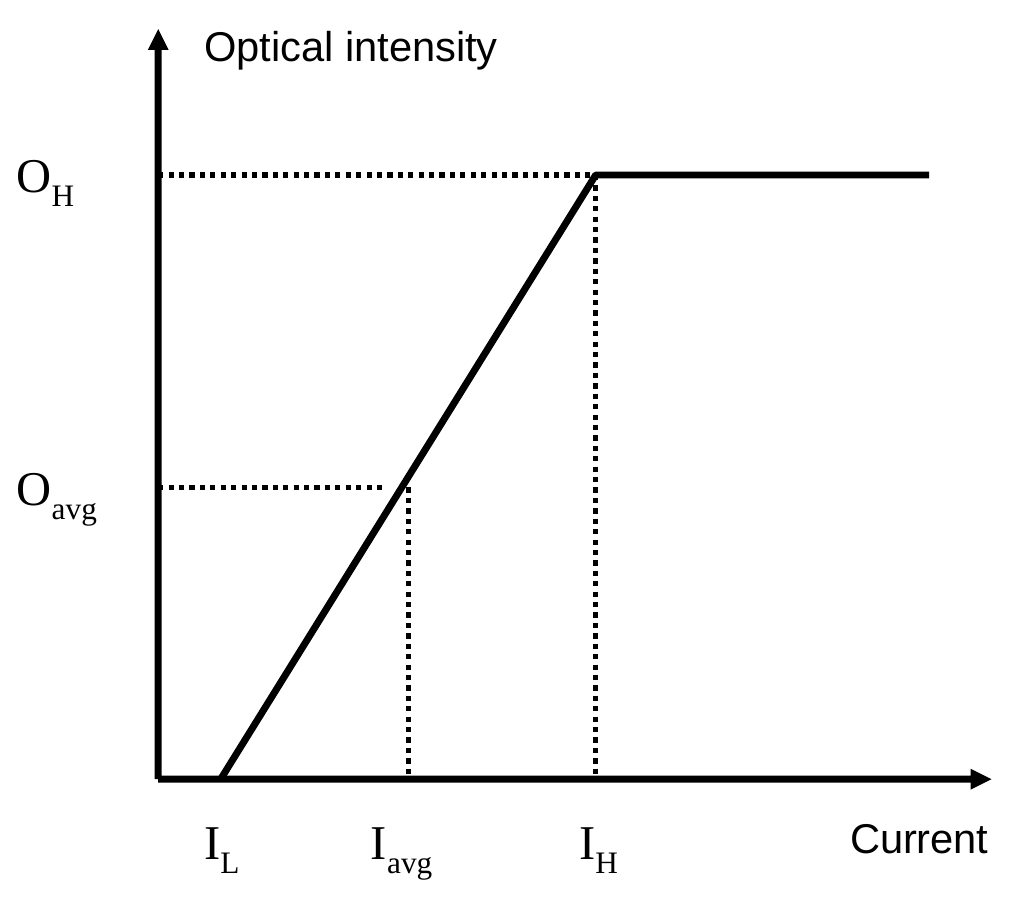}
\caption{Ideal linear LED characteristic.}
\label{fig_led}
\end{figure}

In VLC, LEDs are the main source of non-linearity.  With predistortion, the input-output characteristic of the LED can be linearized, but only within a limited interval $[I_L, I_H]$, where $I_L$ denotes the minimum input current and $I_H$ denotes the maximum input current. The Dynamic range can be denoted by $D \triangleq  I_H-I_L$. Fig.$\,$\ref{fig_led} shows the current-intensity characteristic of an ideal LED. $O_H$ denotes the maximum output amplitude.
The illumination level determines the average output intensity, which is set to a fixed value $O_{\mathrm{avg}}$. Let us denote by $I_{\mathrm{avg}}$ the average input current corresponding to $O_{\mathrm{avg}}$.


IM/DD schemes require the baseband signal in the VLC to be real-valued. To generate real-valued baseband OFDM signal, DC biased optical OFDM (DCO-OFDM) \cite{hranilovic2005design} was introduced for the VLC. According to the property of the inverse Fourier transform, a real-valued
time-domain signal $x(t)$ corresponds to a frequency-domain signal
$X_k$ that is Hermitian symmetric; i.e., $X_k = X_{N-k}^*, 1\leq k \leq N-1$,  
where $*$ denotes complex conjugate. In DCO-OFDM, the $0$th and $N/2$th subcarrier are null; i.e., $X_0 = 0$, $X_{N/2} = 0$. The time-domain signal
$x(t)$ can be obtained from the frequency-domain signal $X_k$ as 
$x(t) =
\frac{1}{\sqrt{N}}\sum_{k=0}^{N-1}X_k\exp(j2\pi kt/T), t \in (0,T]$,
where 
$j = \sqrt{-1}$, and $T$ denotes one OFDM symbol duration. Since the DC component is zero ($X_0 = 0$), $x(t)$ has zero mean. Let us denote by $\sigma_x^2$ the variance of $x(t)$. Let us define the upper peak-to-average power ratio (UPAPR) of $x(t)$ as $\mathcal{U} \triangleq  \left(\underset{t \in (0,T]}{\max} x(t)\right)^2 / \sigma_x^2$, and the lower peak-to-average power ratio (LPAPR) of $x(t)$ as $\mathcal{L} \triangleq  \left(\underset{t \in (0,T]}{\min} x(t)\right)^2 / \sigma_x^2$.

\subsection{Linear scaling and biasing}
\label{}
The forward signal $y(t)$ is obtained from the OFDM signal $x(t)$ after both a linear scaling and a biasing operation; i.e., $y(t) = \alpha x(t) + B, t \in (0,T]$,
where $\alpha$ and $B$ are both real-valued. The resulting signal, $y(t)$, has a mean value $B$ and a variance $\sigma_y^2 = \alpha^2\sigma_x^2$.  The variance $\sigma_y^2$ can be maximized by selecting a scaling factor with the greatest absolute value $|\alpha|$ for each OFDM symbol. To ensure $y(t)$ is within the dynamic range of the LED, we can obtain an $\alpha$ with the greatest absolute value as
\begin{equation}
\label{eq_alphap}
\alpha^{(+)} = \min \left\{ \frac{I_H - B}{\underset{t \in (0,T]}{\max} x(t)}, \frac{I_L - B}{\underset{t \in (0,T]}{\min} x(t)} \right\}, \quad \mathrm{when} \,\,\alpha > 0,
\end{equation}
or 
\begin{equation}
\label{eq_alphan}
\alpha^{(-)} = \max \left\{ \frac{I_H - B}{\underset{t \in (0,T]}{\min} x(t)}, \frac{I_L - B}{\underset{t \in (0,T]}{ \max} x(t)} \right\}, \quad \mathrm{when} \,\,\alpha < 0,
\end{equation}
In other words, an $\alpha$ with the maximum absolute value can be obtained as
\begin{eqnarray}
\label{eq_alphas}
  \alpha = \left\{\begin{array}{cl}
                \alpha^{(+)},  & \mathrm{if} \quad |\alpha^{(+)}| \geq |\alpha^{(-)}| \\
                \alpha^{(-)},  & \mathrm{if} \quad |\alpha^{(+)}| < |\alpha^{(-)}|
              \end{array} \right.
\end{eqnarray}

Let us define the biasing ratio as $\zeta \triangleq (B-I_L)/(I_H-I_L)$. We can obtain the variance of $y(t)$ as
\begin{eqnarray}
\label{eq:var}
\sigma_y^2 
 &=& \sigma_x^2\left(\max \,\, \left\{\alpha^{(+)}, \, -\alpha^{(-)}\right\}\right)^2\\\nonumber
 &=& D^2\max \Bigg\{ \min \left\{\frac{(1-\zeta)^2}{\mathcal{U}}, \frac{\zeta^2}{\mathcal{L}} \right\}, \min \left\{ \frac{(1-\zeta)^2}{\mathcal{L}}, \frac{\zeta^2}{\mathcal{U}} \right\} \Bigg\}.
\end{eqnarray}
\begin{figure}[!t]
  \centering
  \includegraphics[width=6.8cm]{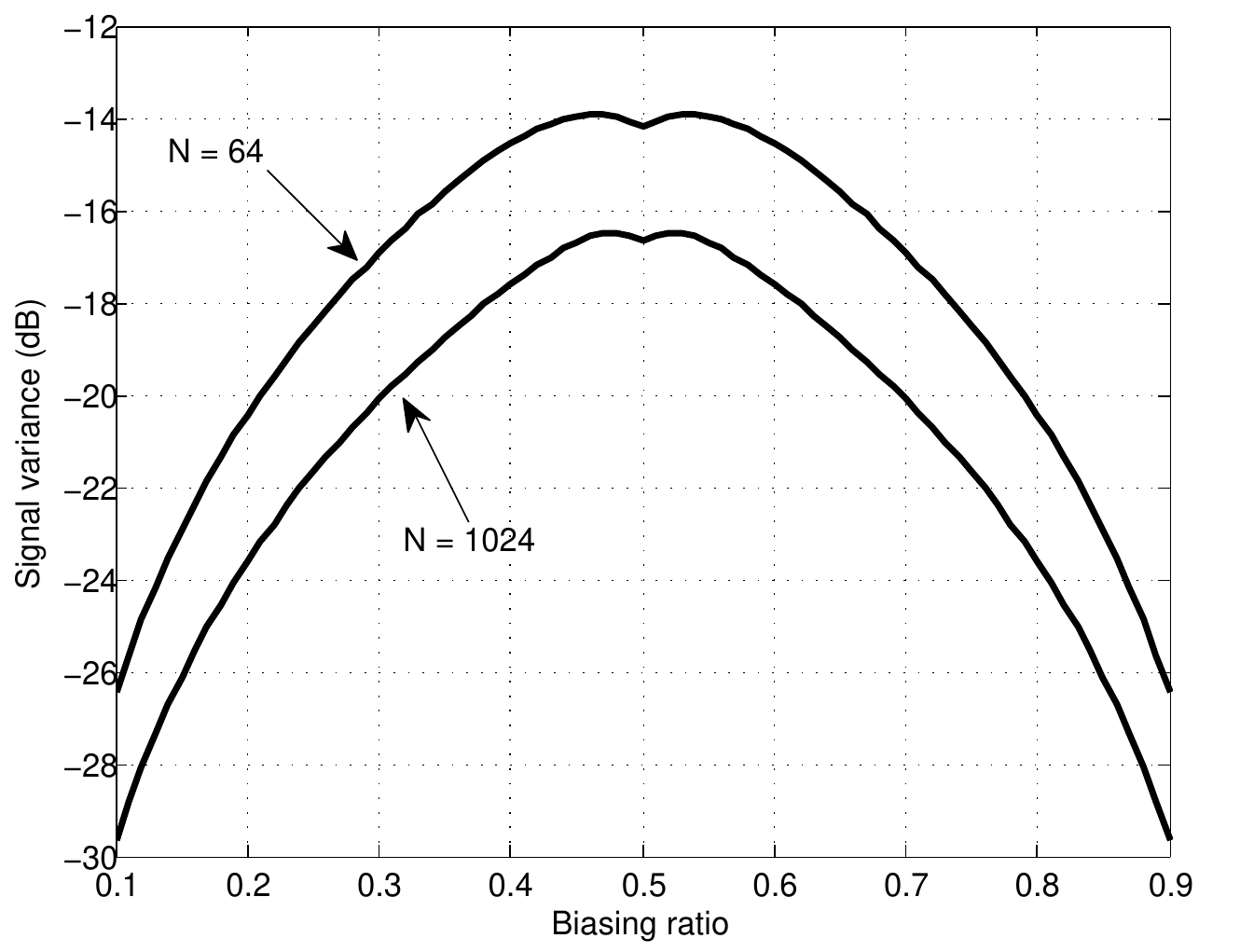}
\caption{Variance $\sigma_y^2$ as a function of the biasing ratio, obtained from 10000 scaled DCO-OFDM symbols with normalized dynamic range.}
\label{fig_var}
\end{figure}
We can observe that the variance $\sigma_y^2$ depends on three factors: biasing ratio, upper PAPR of the OFDM signal and lower PAPR of the OFDM signal. In VLC, the dynamic range $D$ is a fixed value, which is determined by characteristics of LEDs. The scaling factor $\alpha$ varies symbol by symbol since $\mathcal{U}$ and $\mathcal{L}$ are both random variables. We treat $\alpha$ as part of the channel and assume that $\alpha$ for each symbol can be perfectly estimated at the receiver. It has been discussed in reference \cite{jiang2008derivation} that the distribution of PAPR is independent of the constellations but are mainly determined by the number of subcarriers.
Fig. \ref{fig_var} shows the variance $\sigma_y^2$ as a function of the biasing ratio, taken from 10000 scaled DCO-OFDM symbols with normalized dynamic range. The variance decrease with increasing subcarriers because both the UPAPR and LPAPR will increase when there are more subcarriers. Since the DCO-OFDM signal has a symmetric distribution, the maximum variance occurs around biasing ratio 0.5 when the DCO-OFDM signal is biased around the middle point of the dynamic range. However,  the maximum variance occurs at $\zeta = \zeta^{\dagger}$ and
$\zeta = 1 - \zeta^{\dagger}$ symmetrically rather than $\zeta = 0.5$ because $Pr\left(\frac{\mathcal{U}}{\mathcal{L}} = 1\right) < Pr\left(\frac{\mathcal{U}}{\mathcal{L}} = \frac{\zeta^{\dagger}}{1 - \zeta^{\dagger}}\right) + Pr\left(\frac{\mathcal{U}}{\mathcal{L}} = \frac{1 - \zeta^{\dagger}}{\zeta^{\dagger}}\right)$. 

\section{Brightness control}
\label{}
The idea of brightness control is to make the average input current equal to $I_{\mathrm{avg}}$, which corresponds to the desired emitted average intensity $O_{\mathrm{avg}}$. Let us define the brightness factor $\lambda \triangleq O_{\mathrm{avg}}/O_{H} = (I_{\mathrm{avg}}-I_L)/(I_H-I_L)$.
Without loss of generality, we only consider brightness factor in the range $0 \leq \lambda \leq 0.5$, because any forward signal $z(t)$ with  brightness factor $\lambda > 0.5$ can be created from $y(t)$, which has brightness factor $1-\lambda < 0.5$ and is within the dynamic range $[I_L, I_H]$, by $z(t) = I_H+I_L - y(t)$.

We consider two schemes to implement brightness control for DCO-OFDM: (i) biasing adjustment; (ii) pulse width modulation.

\subsection{Biasing adjustment}
Since the mean value of the scaled and biased signal $y(t)$ is equal to $B$, it is straightforward to set the biasing level equal to $I_{\mathrm{avg}}$ for each DCO-OFDM symbol; i.e., $B = I_{\mathrm{avg}}$, $\zeta = \lambda$. Replacing $\zeta$ with $\lambda$ in Eq. (\ref{eq:var}), we can obtain the variance of a scaled DCO-OFDM symbol as
\begin{eqnarray}
\sigma_y^2 = D^2\max \Bigg\{ \min \left\{\frac{(1-\lambda)^2}{\mathcal{U}},\frac{\lambda^2}{\mathcal{L}} \right\},\\\nonumber
\quad\quad\quad\quad \min \left\{ \frac{(1-\lambda)^2}{\mathcal{L}}, \frac{\lambda^2}{\mathcal{U}} \right\} \Bigg\}.
\end{eqnarray}
Define the dynamic-range-to-noise power ratio $\mathrm{DNR} \triangleq G^2D^2/\sigma_N^2$, where $G$ denotes the channel gain and $\sigma_N^2$ denotes variance of the additive white Gaussian noise (AWGN). Thus, the signal to noise ratio (SNR) for each scaled DCO-OFDM symbol can be obtained as
\begin{eqnarray}
\label{eq:snr}
\mathrm{SNR} \triangleq \frac{G^2\sigma_y^2}{\sigma_N^2} = \mathrm{DNR}\cdot \max \Bigg\{ \min \left\{\frac{(1-\lambda)^2}{\mathcal{U}},\frac{\lambda^2}{\mathcal{L}} \right\},\\\nonumber
\quad\quad\quad \min \left\{ \frac{(1-\lambda)^2}{\mathcal{L}}, \frac{\lambda^2}{\mathcal{U}} \right\} \Bigg\}.
\end{eqnarray}
Using the Shannon capacity formula and taking expectation with respect to $\mathcal{U}$ and $\mathcal{L}$, we can obtain the achievable ergodic rates as a function of DNR and $\lambda$ as
\begin{equation}
\mathcal{R}(\mathrm{DNR}, \lambda) = \frac{1}{2}E_{\mathcal{U},\,\mathcal{L}}\left[\log_2(1 + \mathrm{SNR})\right], 
\end{equation}
where the 1/2 degradation is due to the Hermitian symmetry requirement for $X_k$ in the DCO-OFDM system.
\subsection{Pulse width modulation}
PWM is an efficient way to control the brightness of LED. A PWM signal with period $T_{\mathrm{pwm}}$ is expressed as
\begin{equation}
p(t) = 
\begin{cases}
I_{\mathrm{pwm}},& 0 \leq t \leq T\\
0,& T < t \leq T_{\mathrm{pwm}}
\end{cases},
\end{equation}
where $T$ is the ``on'' duration and $T_{\mathrm{pwm}} - T$ is the ``off'' duration. $ I_{\mathrm{pwm}}$ denotes the input current during the ``on'' interval. Let us define the PWM forward ratio $\gamma \triangleq (I_{\mathrm{pwm}} - I_L)/(I_H - I_L).$
The output magnitude can be adjusted by changing the duty cycle $d = T/T_{\mathrm{pwm}}$. To generate the optical intensity with average value $O_{\mathrm{avg}}$, the duty cycle of PWM is chosen to be $d  = (I_{\mathrm{avg}} - I_L)/(I_{\mathrm{pwm}}-I_L) = \lambda/\gamma,$
where $d \leq 1$, $\lambda \leq \gamma$, and $I_{\mathrm{pwm}} \geq I_{\mathrm{avg}}$. 

We propose to combine the DCO-OFDM signal with PWM as
\begin{equation}
y(t) = 
\begin{cases}
\alpha x(t) + I_{\mathrm{pwm}},& 0 \leq t \leq T\\
0,& T < t \leq T_{\mathrm{pwm}}
\end{cases},
\end{equation}
which can be seen as a DCO-OFDM symbol with biasing level $I_{\mathrm{pwm}}$ followed by $T_{\mathrm{pwm}} - T$ length ``0'' compensations.
During the ``on'' interval, the biasing ratio is actually $\gamma$. By replacing $\lambda$ with $\gamma$ in Eq. (\ref{eq:snr}), we can obtain the SNR for each scaled DCO-OFDM symbol during the ``on'' interval as
\begin{eqnarray}
\mathrm{SNR} = \mathrm{DNR}\cdot \max \Bigg\{ \min \left\{\frac{(1-\gamma)^2}{\mathcal{U}},\frac{\gamma^2}{\mathcal{L}}\right\},\\\nonumber
\quad\quad\quad\quad \min \left\{ \frac{(1-\gamma)^2}{\mathcal{L}}, \frac{\gamma^2}{\mathcal{U}} \right\} \Bigg\}.
\end{eqnarray}
Since we do not transmit data in the ``off'' interval, the achievable ergodic rates can be obtained as  
\begin{equation}
\label{eq:rates}
\mathcal{R}(\mathrm{DNR}, \lambda, \gamma) = \frac{1}{2}\cdot\frac{\lambda}{\gamma}\cdot E_{\mathcal{U},\,\mathcal{L}}\left[\log_2(1 + \mathrm{SNR})\right]. 
\end{equation}

In fact, the brightness can be controlled by adjusting the duty cycle or jointly adjusting the duty cycle $d$ and the PWM forward ratio $\gamma$.  Trade-offs exist between $d$ and $\gamma$. Assume the PWM forward ratio falls in a region $\gamma \in [\lambda, \zeta^{\dagger}]$. From the Eq. (\ref{eq:rates}), it can be seen when $\gamma$ is larger, the SNR will increase but the degradation factor $\lambda/\gamma$ will be worse. Therefore, given DNR and brightness factor $\lambda$, we can obtain an optimum $\gamma^*$ that maximize the achievable ergodic rates as $\gamma^* = \underset{\gamma}{\arg \max} \quad \mathcal{R} \,|_{\mathrm{DNR}, \lambda}$.
 
\subsection{Examples}
\begin{figure}[!t]
  \centering
  \includegraphics[width=6.5cm]{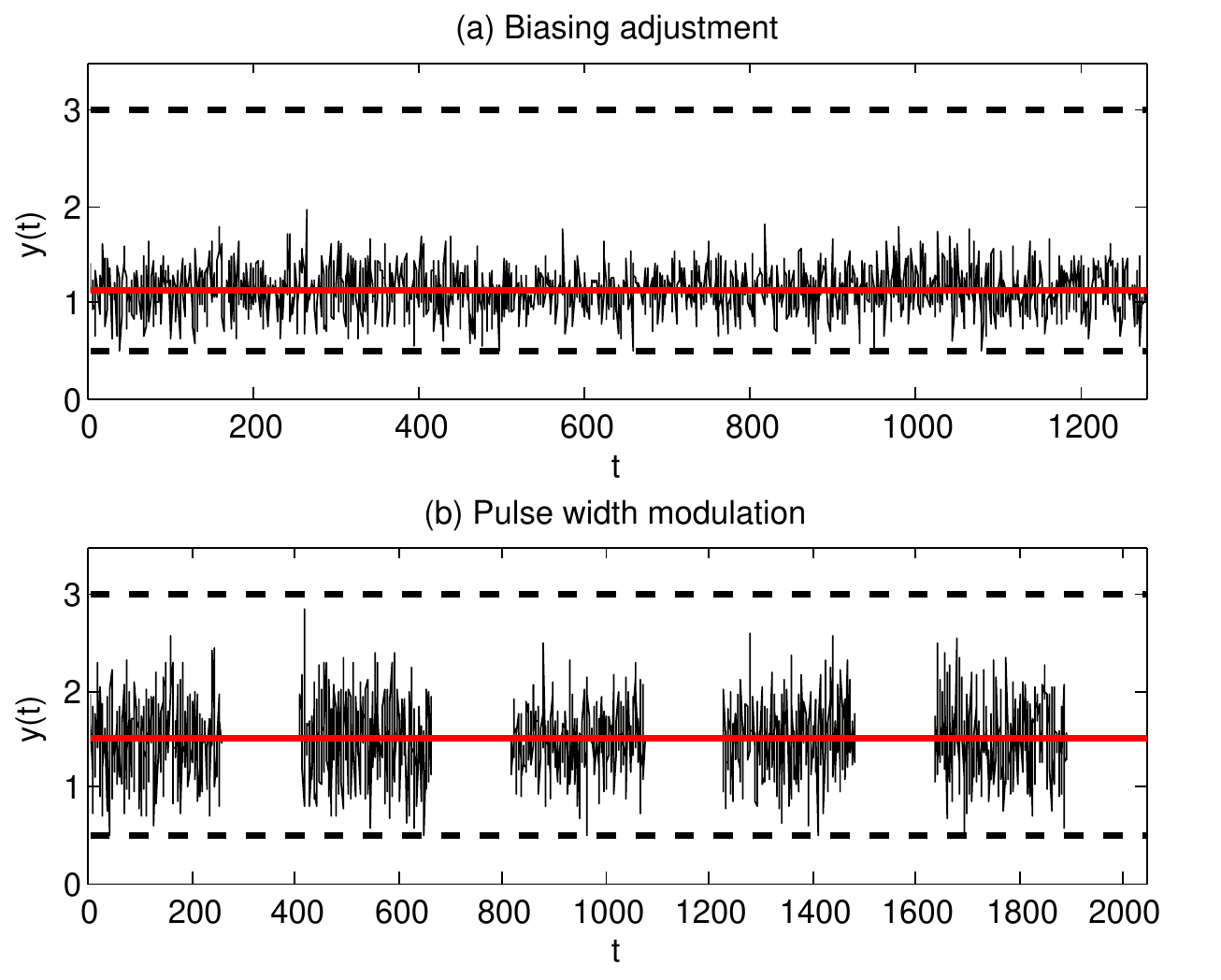}
\caption{An example of transmitting five OFDM symbols with biasing adjustment and PWM schemes (Dash lines: dynamic range of LED; Solid line: biasing level).}
\label{fig_examp}
\end{figure}

To better illustrate the two brightness control schemes under dynamic range constraints, as an example, suppose that we need to transmit five DCO-OFDM symbols with $N= 256$. We assume the brightness factor 
$\lambda$ to be 0.25 and the PWM ratio $\gamma$ is chosen to be 0.4. The corresponding LED input signals $y(t)$ for two schemes are shown in Figure \ref{fig_examp}. 

\section{Numerical results}
\label{sec:num}

\begin{figure}[!t]
  \centering
  \includegraphics[width=7cm]{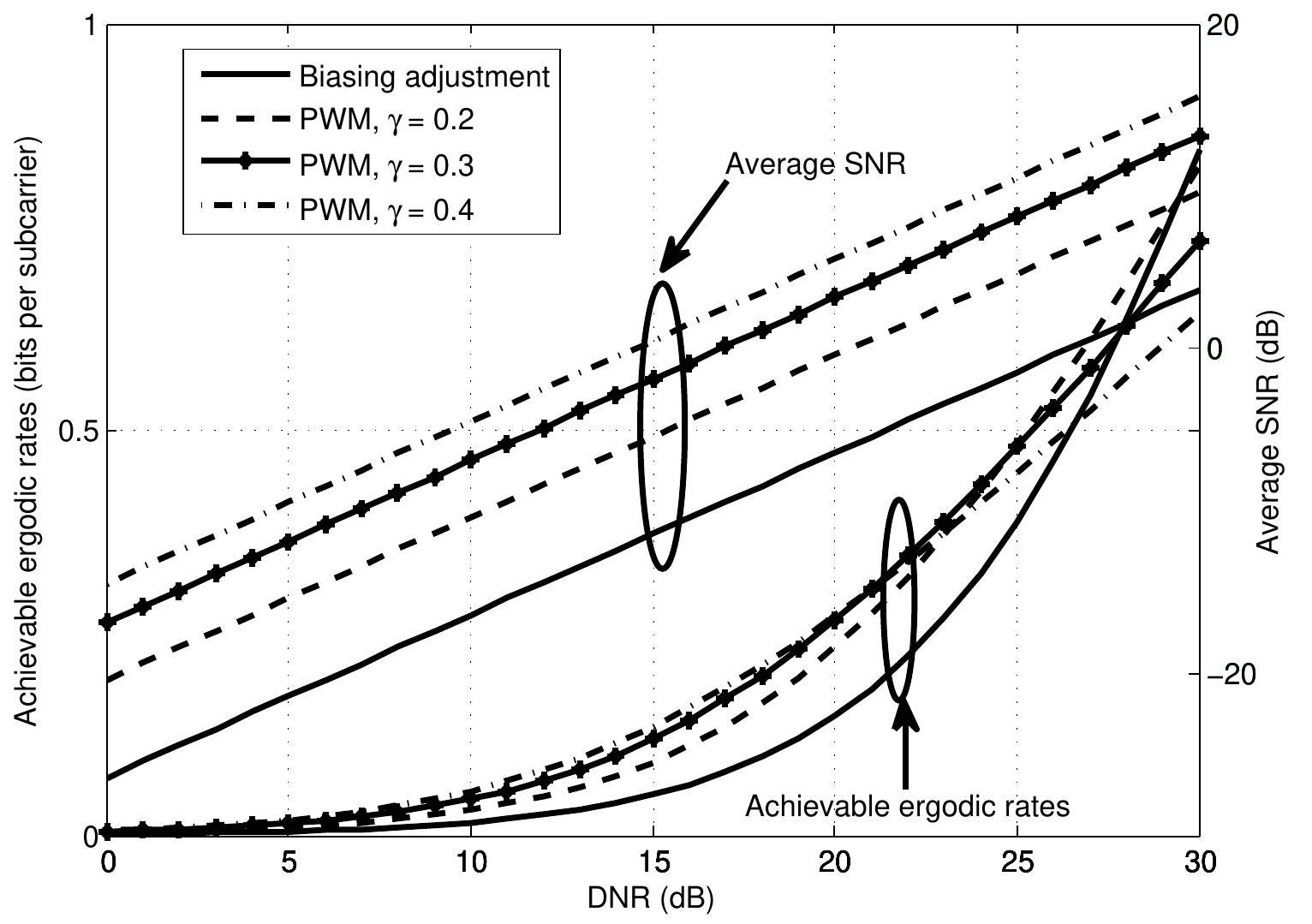}
\caption{Achievable data rates and average SNR as a function of DNR with $\lambda$ = 0.1.}
\label{fig_rate1}
\end{figure}

\begin{figure}[!t]
  \centering
  \includegraphics[width=6.2cm]{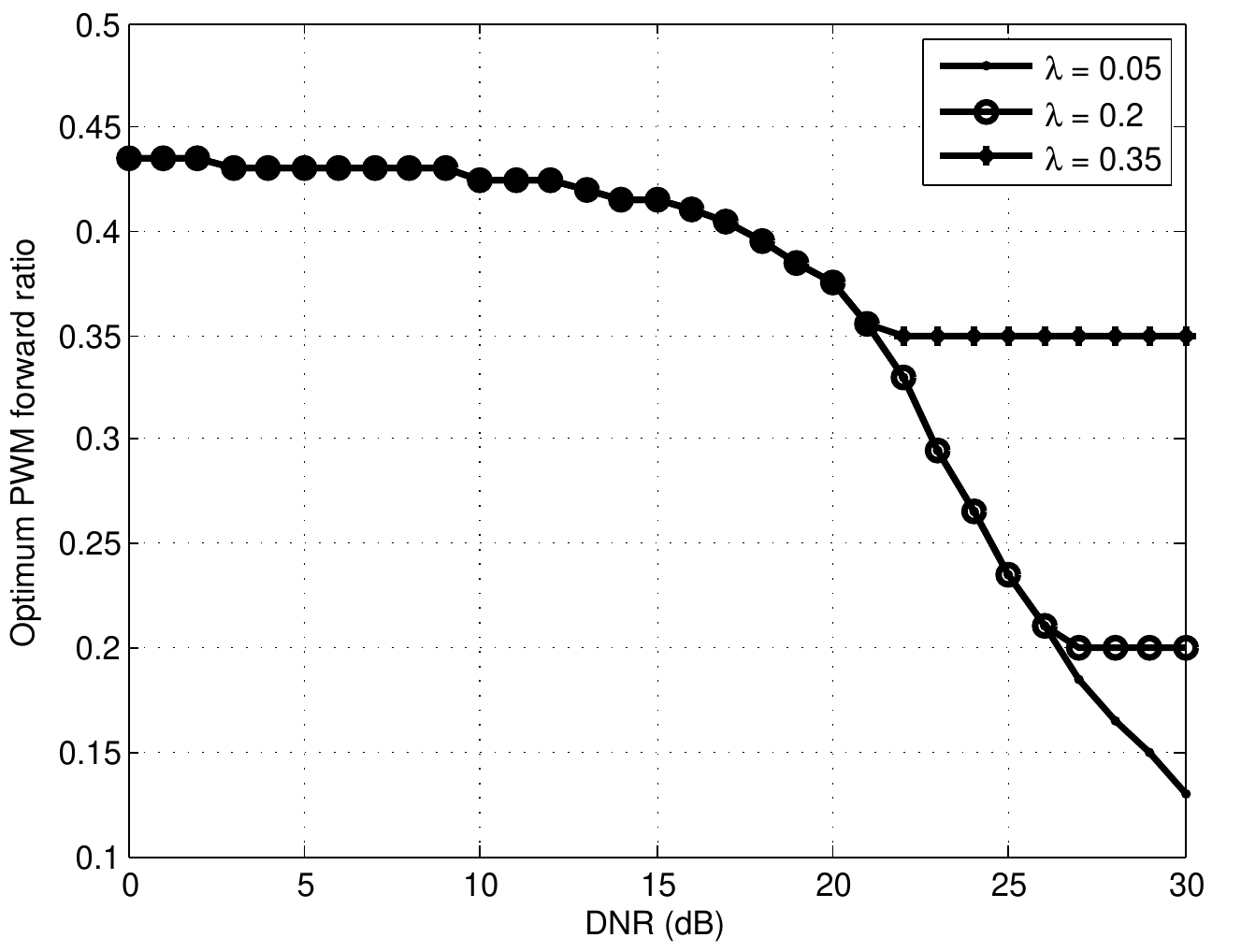}
\caption{Optimum PWM forward ratio as a function of DNR with $\lambda$ = 0.05, 0.2 and 0.35.}
\label{fig_ratio}
\end{figure}

\begin{figure}[!t]
  \centering
  \includegraphics[width=7cm]{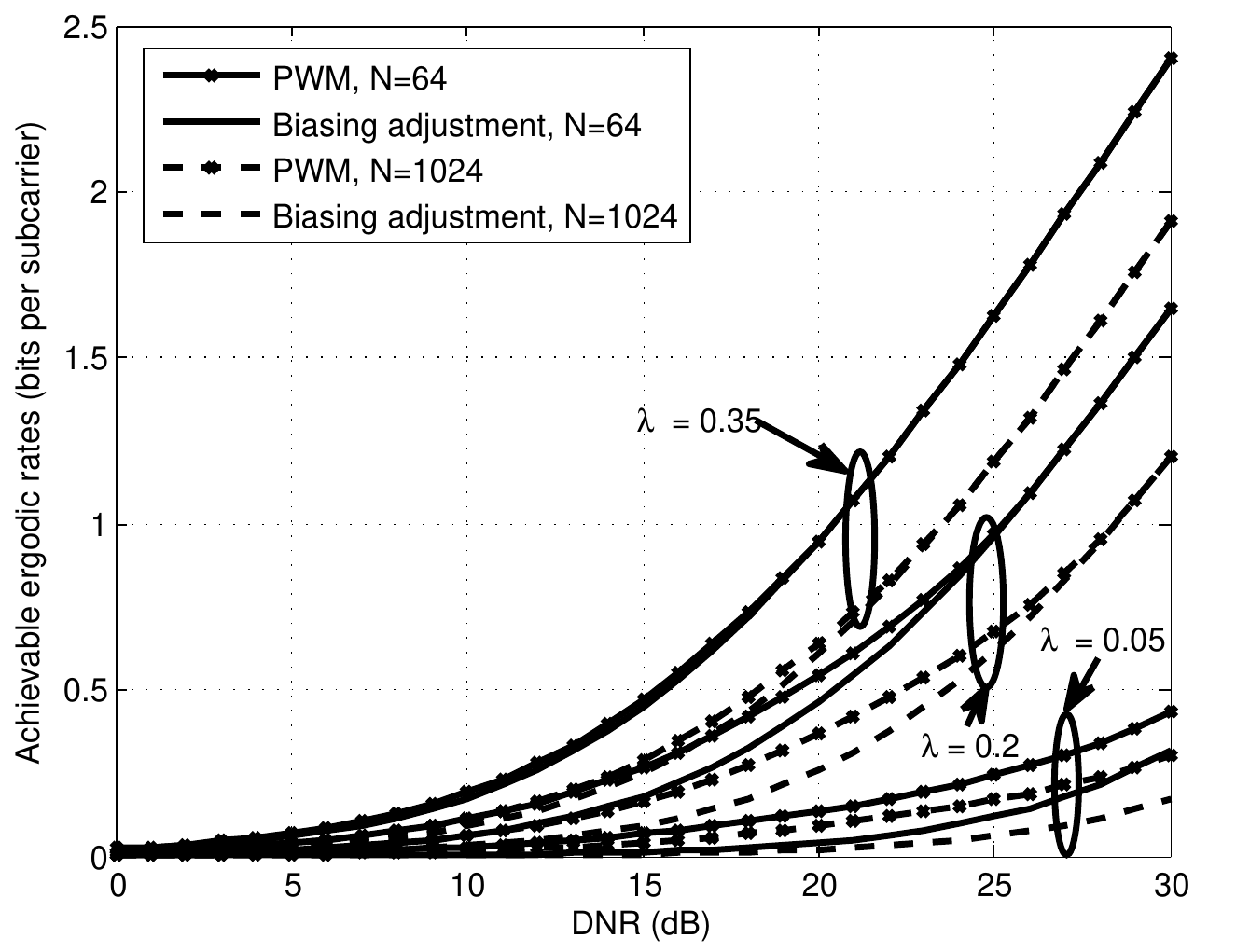}
\caption{Achievable data rates as a function of DNR with optimum PWM forward ratio and $\lambda$ = 0.05, 0.2 and 0.35.}
\label{fig_rate3}
\end{figure}

In this section, we will compare achievable ergodic rates of biasing adjustment method and PWM scheme under various illumination and channel noise scenarios. 
The distribution of UPAPR and LPAPR are drawn from 10000 DCO-OFDM symbols with $N= 64$. Fig. \ref{fig_rate1} shows the achievable ergodic rates and average SNR as a function of DNR with $\lambda$ = 0.1. The PWM ratio $\gamma$ is chosen from 0.2, 0.3 and 0.4. The biasing adjustment method can be seen as a special case of PWM scheme with $d = 1$ and thus $\gamma = \lambda = 0.1$ in that case. 
We can see that although average SNR increase with higher PWM ratio, the achievable ergodic rates depends on the specific $\lambda$ and DNR. Fig. \ref{fig_ratio} shows the optimum PWM forward ratio as a function of DNR with $\lambda$ = 0.05, 0.2 and 0.35. With increasing DNR, the optimum PWM forward ratio $\gamma^*$ approaches $\lambda$. Fig. \ref{fig_rate3} compares the biasing adjusting and PWM with optimum PWM forward ratio with $N=64$ and $N = 1024$. Since biasing adjustment is a special case of the PWM method, PWM with optimum PWM forward ratio will always outperform biasing adjustment. When the brightness factor is larger, the difference become less noticeable.    

\section{Conclusion}

In this letter, we have established a framework for analyzing brightness control in dynamic range constrained visible light OFDM systems. A linear scaling and biasing model was adopted to ensure the forward signal is within the dynamic range of the LED. We have compared the achievable ergodic rates of biasing adjustment and PWM methods for DCO-OFDM. PWM always outperforms biasing adjustment scheme when the optimum PWM forward ratio is chosen. However, the scaling and biasing method may be too ``conservative'' in trying to avoid any distortion and thus not delivering sufficient signal power. An open topic is how to deliberately introduce distortion for performance improvement.


%


%
%

\ifCLASSOPTIONcaptionsoff
  \newpage
\fi



%
\bibliographystyle{IEEEtran}
\bibliography{bmc_article,library}
%

%
%
%




\end{document}